\documentclass{aa}
\usepackage{graphicx}
\usepackage{txfonts}
\usepackage{graphicx}
\usepackage[normalem]{ulem}
\usepackage{floatrow}
\begin{document}

   \title{A  new near-IR  window  of low  extinction  in the  Galactic
     plane\thanks{Based  on observations  taken within  the ESO  VISTA
       Public Survey VVV, programme ID 179.B-2002}}
   \titlerunning{A  window  to  the  far  side  of  the  Galaxy}
   \authorrunning{Minniti et al.}

   %\subtitle{}

\author{ 
Dante Minniti\inst{\ref{inst1},\ref{inst2},\ref{inst3}}\and
Roberto K. Saito\inst{\ref{inst4}}\and
Oscar A. Gonzalez\inst{\ref{inst5}}\and 
Javier Alonso-Garc\'ia\inst{\ref{inst6},\ref{inst2}}\and 
Marina Rejkuba\inst{\ref{inst7},\ref{inst8}}\and
Rodolfo Barb\'a\inst{\ref{inst9}}\and
Mike Irwin\inst{\ref{inst10}}\and
Roberto Kammers\inst{\ref{inst4}}\and
Phillip W. Lucas\inst{\ref{inst11}}\and
Daniel Majaess\inst{\ref{inst12},\ref{inst13}}\and
Elena Valenti\inst{\ref{inst7}}}

\institute{
Departamento  de  Fisica,  Facultad  de  Ciencias  Exactas,
  Universidad  Andres Bello,  Av.  Fernandez  Concha 700,  Las Condes,
  Santiago, Chile\label{inst1}
\and
Instituto Milenio de Astrofisica, Santiago, Chile\label{inst2}
\and
Vatican Observatory, V00120 Vatican City State, Italy\label{inst3}
\and
Departamento de  F\'{i}sica, Universidade  Federal de  Santa Catarina,
Trindade 88040-900, Florian\'opolis, SC, Brazil\label{inst4}
\and
UK Astronomy Technology Centre, Royal Observatory, Blackford Hill,
Edinburgh EH9 3HJ, UK\label{inst5}
\and 
Unidad   de  Astronomia,   Facultad  Cs.    Basicas,  Universidad   de
Antofagasta,   Avda.    U.    de   Antofagasta   02800,   Antofagasta,
Chile\label{inst6}
\and
European  Southern  Observatory,   Karl-Schwarszchild-Str.  2,  D85748
Garching bei Muenchen, Germany\label{inst7}
\and
Excellence Cluster  Universe, Boltzmann-Str.   2, D85748  Garching bei
Muenchen, Germany\label{inst8}
\and
Departamento  de  F\'{\i}sica  y  Astronom\'{\i}a,  University  of  La
Serena, La Serena, Chile\label{inst9}
\and
Department of Astronomy, Cambridge University, Cambridge,
UK\label{inst10}
\and
Department of  Astronomy, University of  Hertfordshire, Hertfordshire,
UK\label{inst11}
\and
Mount    Saint    Vincent    University,   Halifax,    Nova    Scotia,
Canada\label{inst12}
\and
Saint Mary's University, Halifax, Nova Scotia, Canada\label{inst13}
\and
Pontificia Universidad  Catolica de  Chile, Instituto  de Astrofisica,
Av. Vicu\~na Mackenna 4860, Santiago, Chile\label{inst14}
}

  \date{Received XXX XX, 2017; accepted XXX XX, 2017}

  \abstract

\abstract {}  {The windows  of low  extinction in  the Milky  Way (MW)
  plane  are  rare but  important  because  they  enable us  to  place
  structural constraints on the opposite side of the Galaxy, which has
  hitherto  been done  rarely.}  {We  use the  near-infrared (near-IR)
  images of the VISTA Variables  in the V\'{\i}a L\'actea (VVV) Survey
  to  build extinction  maps and  to identify  low extinction  windows
  towards the Southern  Galactic plane.  Here we  report the discovery
  of VVV  WIN 1713$-$3939, a  very interesting window  with relatively
  uniform and  low extinction  conveniently placed  very close  to the
  Galactic plane.}  {The  new window of roughly 30  arcmin diameter is
  located at  Galactic coordinates $(l,  ~b)= (347.4, -0.4)$  deg.  We
  analyse the  VVV near-IR  colour-magnitude diagrams in  this window.
  The mean total near-IR extinction  and reddening values measured for
  this window  are $A_{Ks}=0.46$  and $E(J-K_{\rm s})=0.95$.   The red
  clump giants within the window show a bimodal magnitude distribution
  in the $K_{\rm  s}$~band, with peaks at $K_{\rm  s}=14.1$ and $14.8$
  mag,  corresponding   to  mean   distances  of   $D=11.0\pm2.4$  and
  $14.8\pm3.6$ kpc, respectively.  We discuss  the origin of these red
  clump overdensities  within the context  of the MW  disk structure.}
          {}


   \maketitle

\section{Introduction} 
\label{sec:intro}

\begin{figure*}
\floatbox[{\capbeside\thisfloatsetup{capbesideposition={right,top},capbesidewidth=4cm}}]{figure}[\FBwidth]
         {\caption{Extinction maps for the inner Galactic plane.  Top:
             The  AKS extinction  map,  produced using  the method  of
             Majewski   et   al.     (2011),   covering   $342<l<352$,
             $-2<b<+2$~deg.  Middle: The EJKS  map, produced using the
             method of  Gonzalez et  al.  (2011)  for the  same region
             (see  Section   2).   Horizontal  lines  are   at  $b=0$,
             indicating the Galactic plane, and at $|b|=0.8$. Circular
             areas mark the  region around the window  and the control
             field (see  Section 2). In  the bottom panel is  the EJKS
             map  zoomed  in the  window's  region.   The squared  and
             circular  areas   marked  in   white  are   described  in
             Figs.~\ref{fig:map2} and  \ref{fig:gaia}.  The horizontal
             solid line  indicates the  Galactic plane.   The vertical
             bar shows the colour code for each map.}
\label{fig:map}}
{\includegraphics[scale=0.75]{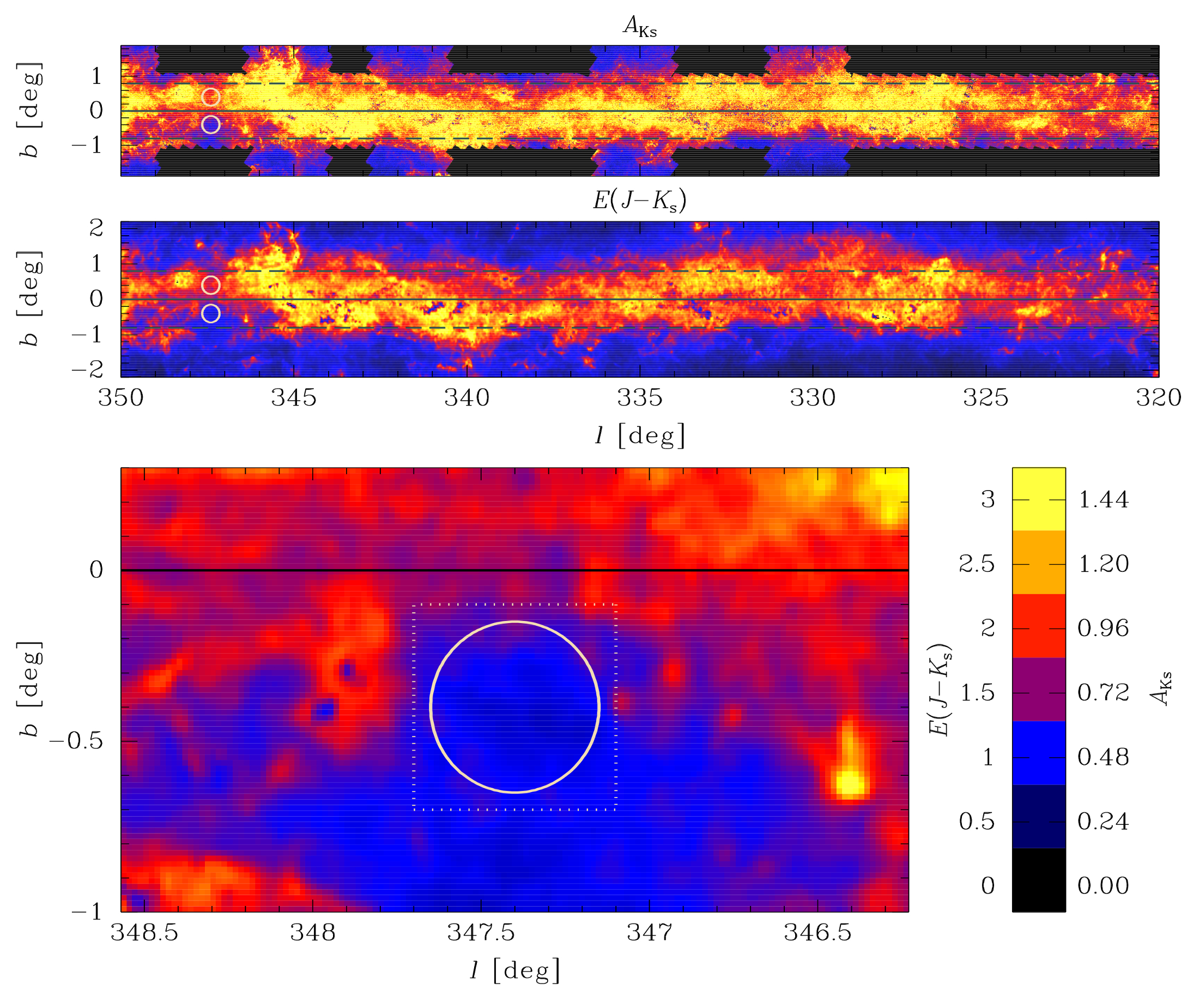}}
\end{figure*}

Baade's window  (Baade \&  Gaposchkin 1963) is  a low  extinction $A_V
\sim 1.5$ region (Schlegel et al.  1998, Schlafly et al. 2011) of half
a  degree  in size  in  the  direction of  the  Milky  Way (MW)  bulge
$(l,~b)=(+1^{\circ},-4^{\circ})$.  This  region has  historically been
used  as  a  fundamental  reference  for  the  study  of  the  stellar
populations  of  the  Galactic  bulge.  The  search  for  similar  low
extinction  regions,  particularly  at  low latitudes  closer  to  the
Galactic plane  ($|b| < 1^\circ$~deg),  is an important task  that can
facilitate  precise  studies  of  the stellar  populations  at  larger
distances along the line of  sight, whilst minimising the difficulties
arising from differential extinction and completeness.

Large   and  non-uniform   reddening  confuses   studies  of   stellar
populations in the Galactic  plane, resulting sometimes in conflicting
conclusions   with   regard   to   the   structure   of   our   Galaxy
(e.g. Vall\'{e}e 2014, 2017,  and references therein).  Additional low
latitude windows  with low extinction can  be useful for a  variety of
reasons especially at shorter wavelengths (as was the case for Baade's
window in the bulge), such as  to verify the next generation models of
star  counts  in   the  MW  disk,  to  identify   distant  sources  in
multi-wavelength studies and to  enable their spectroscopic follow-up,
to use as  a calibration for other more reddened  fields, to check the
behaviour  of the  extinction  law along  the line  of  sight, and  to
measure the total transparency of our Galaxy.
 
Existing extinction maps  (e.g. Schlegel et al. 1998,  Schlafly et al.
2011) show  a few  places of  low extinction  near the  inner Galactic
plane  located at  latitudes $1^{\circ}  <|b| <  2^{\circ}$.  However,
given  the thin-disk  scale  height  of $h_z=$  300  pc (Juri\'{c}  et
al. 2008),  a line  of sight towards  $|b| > 1$  deg departs  from the
plane at large distances ($\rm> 20 kpc$) and thus is not as useful for
mapping the far side of the MW disk.

The VISTA Variables in V\'{\i}a L\'actea (VVV) near-infrared (near-IR)
imaging survey has  recently completed mapping of  a 562~sq.~deg. area
of the MW  bulge and the adjacent plane with  the Visible and Infrared
Survey  Telescope  for Astronomy  (VISTA),  operated  by the  European
Southern Observatory (ESO) at the  Cerro Paranal Observatory in Chile.
The whole area  was imaged twice in $ZYJHK_{\rm s}$  filters, and more
than 50 times in $K_{\rm s}$~band  (Minniti et al.  2010, Saito et al.
2012), providing  the deepest  and highest spatial  resolution near-IR
data set available for the study of the inner Milky Way.

The distribution of red clump  (RC) stars (core helium burning giants)
in  the  VVV data  has  been  used to  trace  the  inner Galactic  bar
(Gonzalez et al.   2011a) and the boxy or peanut  shape bulge (Wegg \&
Gerhard 2013), to map the edge of  the MW stellar disk (Minniti et al.
2011), to study the interstellar extinction law at low latitudes (e.g.
Nataf et al.  2016, Majaess et  al.  2016, Alonso-Garcia et al. 2017),
and to  make detailed maps  of these  regions to study  the extinction
spatial distribution (Gonzalez et al.   2011b, 2012, Schultheis et al.
2014,  Soto et  al.   2013,  Minniti et  al.   2014).  Recently,  when
exploring  the VVV  density  and  extinction maps,  we  found VVV  WIN
1713$-$3939, a new  window with low and  relatively uniform extinction
at $(l, ~b)=  (347.4^{\circ}, -0.4^{\circ})$, which is  the subject of
the present study.

\section{A new Galactic window}
\label{sec:sec2}

We performed a  search for low latitude, low-reddening  windows in the
Galactic  plane  using the  multi-colour  $ZYJHK_{\rm  s}$ images  and
photometry   of   the   disk   and  bulge   fields   in   the   region
($-2.25^{\circ}<b<2.25^{\circ}$,  $295^{\circ}<l<10^{\circ}$) observed
by the VVV survey  (Saito et al. 2012). We used  two different sets of
reddening maps that have been  independently made. The first reddening
map was constructed following the  method described in Gonzalez et al.
(2011, 2012), and computed in  units of colour excess $E(J-K_{\rm S})$
(hereafter  called  ``EJKS''  map).   The   latter  is  based  on  the
Rayleigh-Jeans  colour excess  method  (e.g.  Majewski  et al.   2011,
Nidever et  al.  2012) and computed  in terms of the  total extinction
$A_{Ks}$  (hereafter ``AKS''  map).  Visual  inspection of  these maps
reveals a window of considerably  lower extinction with respect to the
neighbouring regions located towards the inner regions of the MW plane
at     $(l,    ~b)     =    (347.4^{\circ},     -0.4^{\circ})$    (see
Fig.~\ref{fig:map}). This  is the only  large region of  uniformly low
extinction at $|b|<0.8$ seen along across all of the maps.

This window can  also be seen in the Spitzer  Galactic Legacy Infrared
Mid-Plane Survey  Extraordinaire (GLIMPSE) survey images  (Benjamin et
al.  2003, Churchwell et al.  2009), in the Planck R2-HFI maps (Planck
collaboration 2011), in the  CO maps of Dame et al.  (2001) as well as
in the density maps of Nidever et al.  (2012) and Soto et al.  (2013).
In all cases,  the window can be identified not  only in $A_K$ values,
but also as an enhancement in the number of detected stars.  A similar
increment in the source density is  also seen in the Gaia optical data
(Gaia Collaboration et al.  2016).

\begin{figure}
\centering
\includegraphics[scale=0.53]{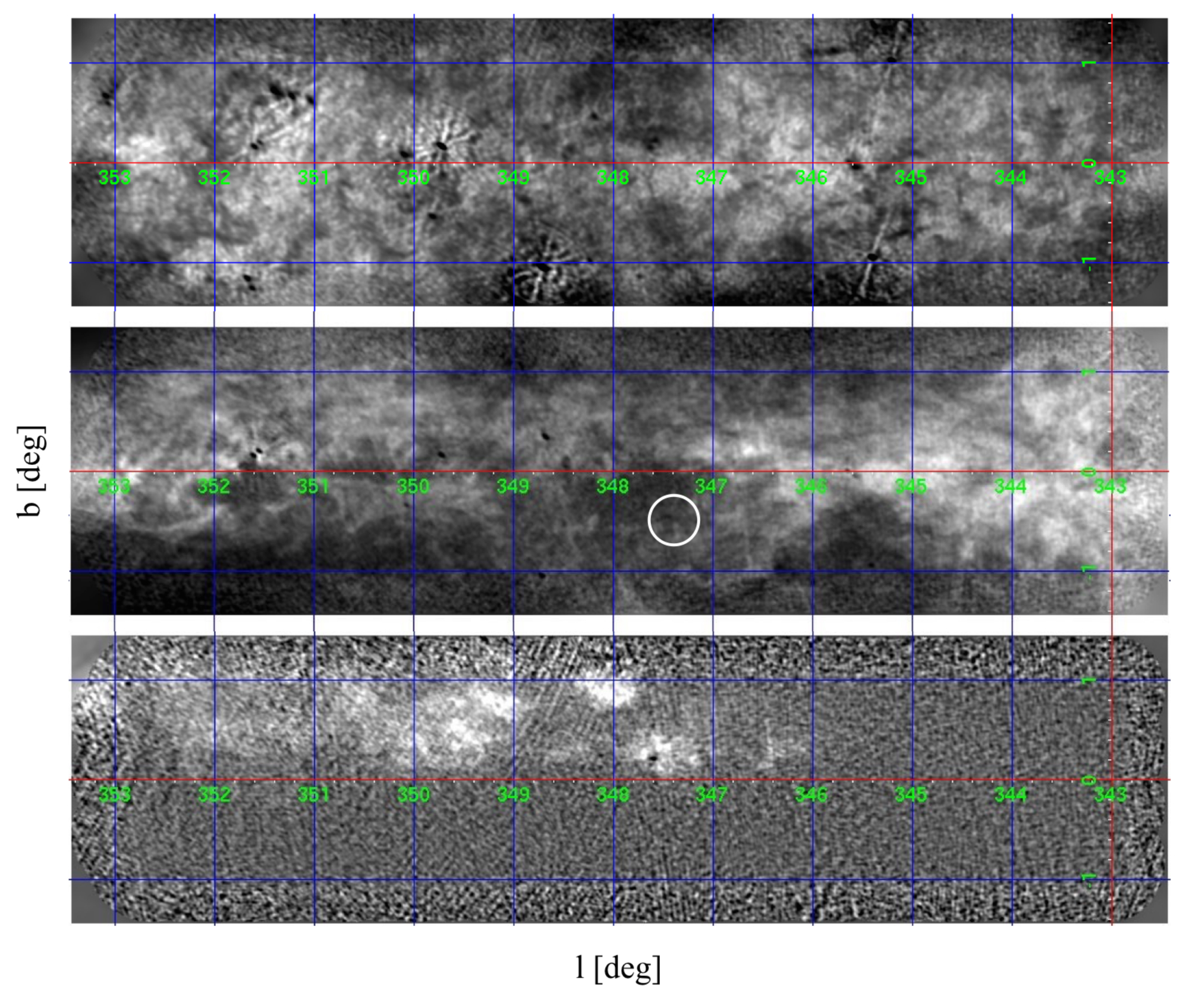}
\caption{HI maps at different velocities for the inner Galactic plane,
  for  $342.6<l<353.4$   and  $-1.4<b<1.4$.   Top:  HI   emission  for
  velocities between $-15$ to $-20$~km/s.  Middle: Same for velocities
  between  $-30$ to  $-42$~km/s. The  location of  the low  extinction
  window is  indicated with the  circle.  Bottom: Same  for velocities
  between $-150$ to $-185$~km/s.}
\label{fig:HI}
\end{figure}

\begin{figure}
\centering
\includegraphics[scale=0.6]{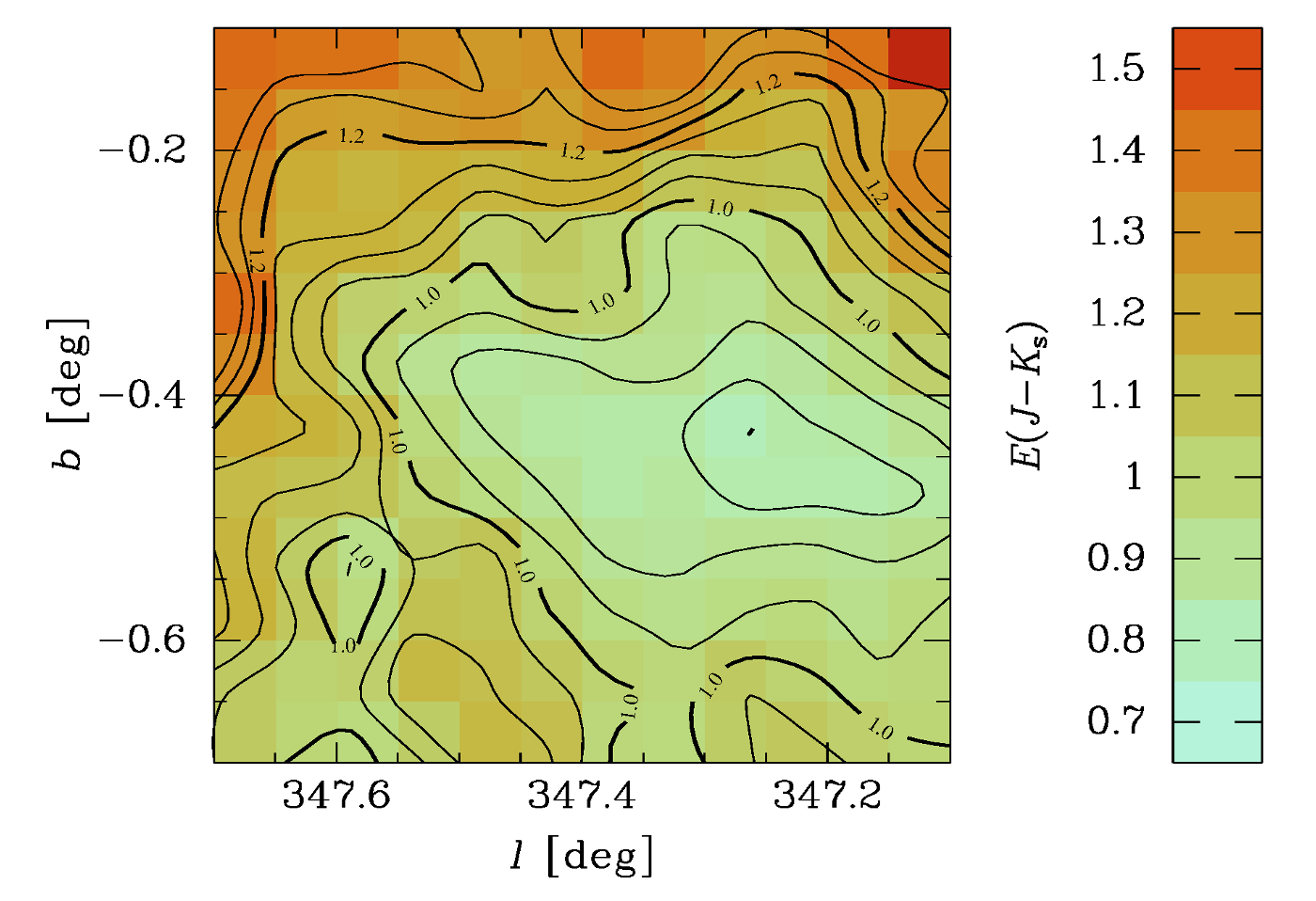}
\caption{Detailed view of VVV~WIN~1713$-$3939.  The map is $36' \times
  36'$  in size,  has a  resolution of  $3'$, and  corresponds to  the
  squared region marked in Fig.~\ref{fig:map}.  The contour lines show
  levels of similar $E(J-K_{\rm S})$, in steps of 0.05 mag.}
\label{fig:map2}
\end{figure}

VVV~WIN~1713$-$3939  has a  roughly circular  shape of  $30'$ diameter
size, centred  at $(l,  ~b) =  (347.4^{\circ}, -0.4^{\circ})$  deg.  A
detailed    view    of    VVV~WIN~1713$-$3939    is    presented    in
Fig.~\ref{fig:map2}.   For a  Galactic latitude  of $b=-0.4$  deg, the
projected  vertical height  of the  line of  sight below  the Galactic
plane at a distance of $D=8$ kpc corresponds to $z = -56$ pc, and at a
distance of $D=14$  kpc, this height is  $z = -98$ pc,  thus making it
ideal to  trace the properties  of the  thin disk beyond  the Galactic
bar.

Reading  from the  AKS extinction  maps,  the mean  extinction in  the
window (integrated to a distance  of $\sim$10 kpc) is $A_{Ks}=0.62 \pm
0.07$.  Reading from the EJKS map, the corresponding mean reddening of
the window is  $E(J-K_{\rm s})=0.96 \pm 0.09$, which  is equivalent to
$A_{Ks}=0.46 \pm 0.04$.  In general, both maps are  in good agreement,
however the AKS maps tend to show higher values than EJKS.  A pixel to
pixel division (AKS/EJKS map, both in  units of $A_{Ks}$) results in a
median value of $1.04 \pm  0.37$ for $A_{Ks}<1.2$, with larger scatter
for increasing $A_{Ks}$ values.

A comparison with the maps of  Schlegel et al.  (1998) and Schlafly et
al.   (2011) reveals  that they  obtain  larger values  for the  total
extinction in this direction,  $A_{Ks}= 1.4$ and $1.6$, respectively.
This  systematic difference,  in the  sense that  the $A_{Ks}$  in the
near-IR is generally  lower than the $A_{Ks}$ measured  in the far-IR,
is  seen throughout  the  Galactic  plane regions  mapped  by the  VVV
survey.

\begin{figure}
\centering
\includegraphics[scale=0.5]{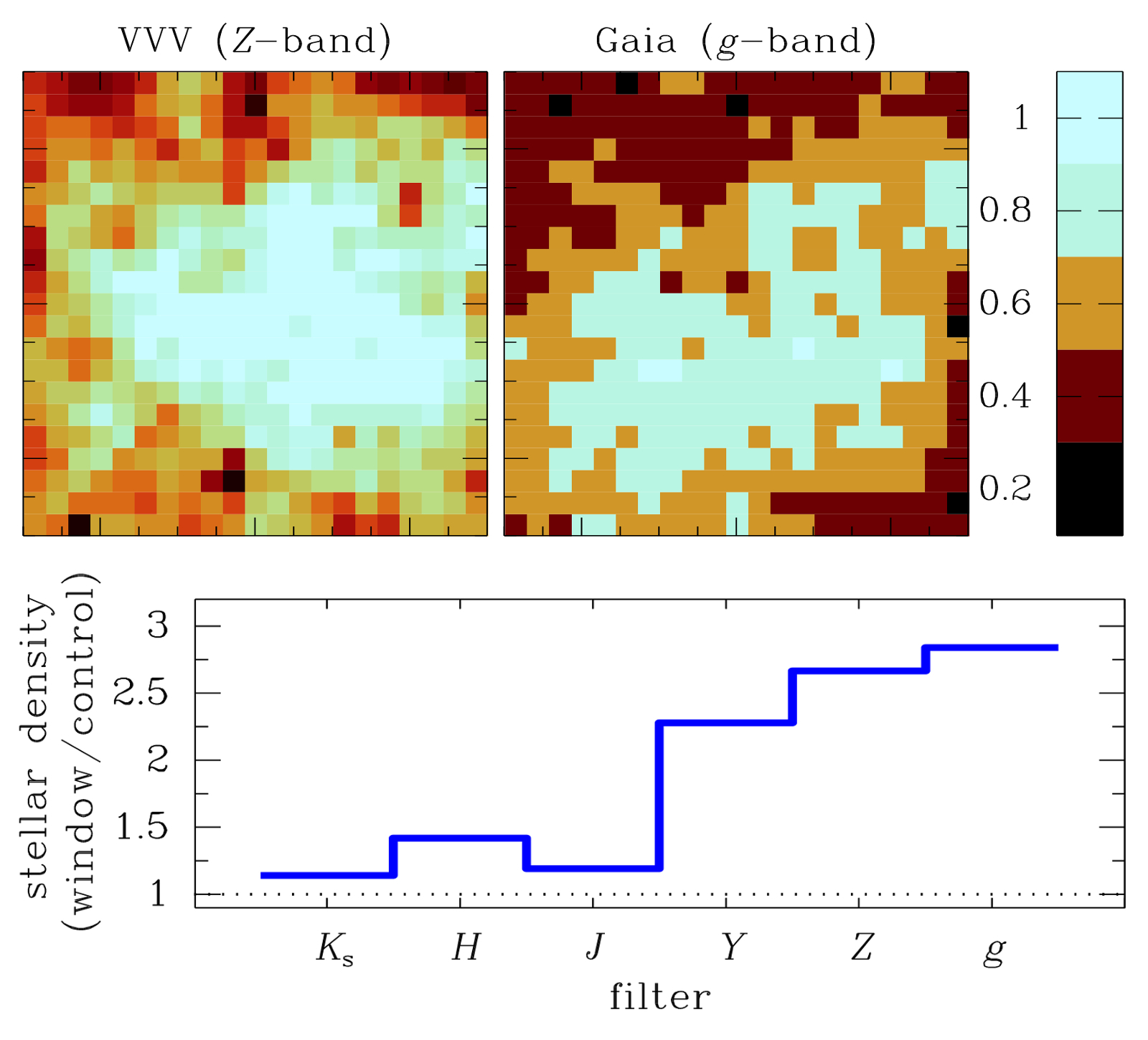}
\caption{Top: The  same region of  Fig.~\ref{fig:map2} as seen  in the
  density maps  of VVV ($Z$~band)  and Gaia ($g$~band),  normalised by
  the maximum value in each case. Vertical bar shows the colour code in
  the  maps. Bottom:  Total number  of sources  in the  low extinction
  window divided by the number of sources in the control field for the
  VVV filters and $g$~band from Gaia.}
\label{fig:gaia}
\end{figure}

In  order  to  interpret   the  colour-magnitude  diagrams  (CMDs)  of
VVV~WIN~1713$-$3939 we compare it with  a control field located at the
same  Galactic  coordinates,  but  at the  positive  latitudes,  i.e.,
centred at  $(l, ~b)  = (347.4,  +0.4)$ deg  (see Fig.~\ref{fig:map}).
The Galactic  warp is negligible  at these longitudes (e.g.  Momany et
al.   2006, see  their  Fig.   9), thus  the  differences between  the
control  field  and  the  window  are  expected  to  be  dominated  by
extinction variation along the line of sight ($\Delta K_{\rm s} = 0.5$
mag).  An  examination of the density  maps shows a higher  density of
sources within  the window  in all  VVV filters  when compared  to the
control field.   For instance,  in $Z$~band which  is the  filter most
affected by extinction, the  stellar density in VVV~WIN~1713$-$3939 is
2.7 times larger  than that of the control field.   There is a similar
difference in stellar density observed  in the optical Gaia data (Gaia
Collaboration  et al.   2016).   The Gaia  source  density within  the
window area is  three times larger than the control  field.  Also, the
spatial distribution  of the  stellar overdensity seen  in Gaia  is in
agreement with the  location of the window seen in  the extinction and
density maps from VVV (see Figs.~\ref{fig:map2} and \ref{fig:gaia}).

Figure~\ref{fig:HI}  shows  as  example  three HI  maps  at  different
velocities,  illustrating   the  relative   absence  of   emission  in
VVV~WIN~1713$-$3939.  The window in HI suggests a chance superposition
of different  sizable holes in  the near and  far side of  the tangent
point.  At the tangent point in $l=347.4$~deg ($-185$~km/s), where the
HI density  is expected to  be higher, most of  the gas appears  to be
located above the plane with  $|b|>0$~deg.  The more recent search for
Galactic HII regions using  Wide-field Infrared Survey Explorer (WISE)
data also  shows a lack of  sources in the low  extinction window, and
confirms that  most known  and new candidate  HII regions  are located
above $b=0$ deg  for this Galactic longitude (Anderson,  et al. 2014).
However, lack  of HI on its  own does not necessarily  mean absence of
dust.  We note that in this window  there is also a lack of integrated
CO   (1---0   emission   line   at   115   GHz)   in   the   maps   of
\cite{2001ApJ...547..792D}.     For    the    future,    a    detailed
three-dimensional (3-D) comparison with all  dust tracers (HI, CO, HII
and dark gas) along the line of sight to this window is needed.

\section{The distribution of red clump stars through VVV~WIN~1713$-$3939} 
\label{sec:sec3}

Once we have identified the low  extinction window, we can explore its
stellar content  along the line of  sight, using RC stars  as distance
indicators.  RC  stars are  core helium  burning giants  and excellent
distance  indicators,  and  have  been used  in  multiple  studies  of
Galactic  structure, in  particular  those aimed  at  tracing the  bar
(e.g. Stanek et  al.  1994; Gonzalez et al.  2011)  and the structures
towards  the inner  MW  bulge  (Nishiyama et  al.   2005, Gonzalez  et
al. 2012).  Also,  Benjamin et al.  (2005) devised a  method to use RC
giants from  the GLIMPSE  survey to investigate  the structure  of the
Galactic plane.  This method works well when using GLIMPSE data out to
the distance of about 8 kpc  (Benjamin et al.  2005), enabling mapping
of the near-end  of the Galactic bar and nearby  spiral arms. However,
beyond 8  kpc the  GLIMPSE photometry is  incomplete (see  for example
Fig.  5  of Nidever  et al.   2012).  Alternative  distance indicators
like variable stars  such as RR Lyrae or Cepheids  are interesting but
of  limited use  in  this  case.  The  RR  Lyrae  stars represent  old
metal-poor populations,  which are very rare  if not absent in  the MW
disk.   In  the  case  of  the Cepheids,  they  represent  very  young
populations, but their surface density is low.

\begin{figure}
\centering
\includegraphics[scale=0.68]{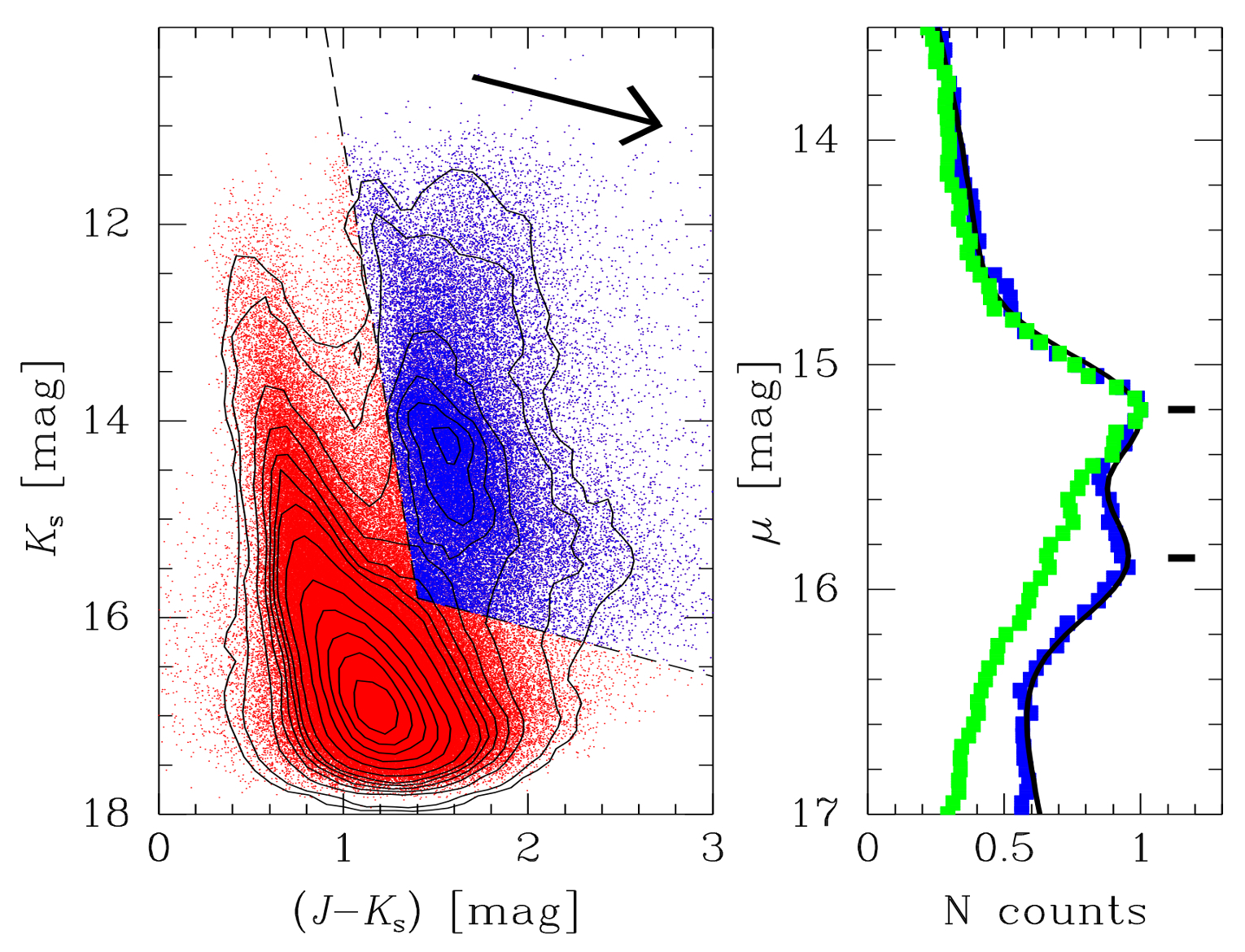}
\caption{Left: $K_{\rm s}$~versus~$(J-K_{\rm s})$ CMD for the circular
  region around the  window as shown in  Fig.~\ref{fig:map}.  The blue
  region was  used to derive  the RC  mags and colours.   The contours
  trace  iso density  levels. A  reddening  vector with  the slope  of
  $A_{Ks}=0.484  \times   E(J-K_{\rm  s})$  is  also   shown.   Right:
  Normalised  density  distribution  of selected  RC  versus  distance
  moduli  of the  low  extinction  window (blue  squares)  and of  the
  comparison field (green squares).}
\label{fig:cmd}
\end{figure}

We produce the CMD of  VVV~WIN~1713$-$3939 and the control field using
the VVV point spread function  (PSF) catalogues described in detail in
Alonso-Garc\'ia et al.  (2015, 2017).  One of the immediate advantages
of a low extinction window is that it is possible to first inspect the
observed  CMD,  minimising  the  effects  from  different  assumptions
regarding the  extinction law in  a de-reddened CMD.  The  $K_{\rm s}$
versus $(J-K_{\rm  s})$ CMD for  the $30'$ diameter window  centred at
$(l, ~b) =  (347.4^{\circ}, -0.4^{\circ})$ is shown in  the left panel
of Fig.~\ref{fig:cmd}.   The observed CMD shows  a well-defined double
peaked RC  feature. We  suggest that these  are indeed  two structures
along the line of  sight. If this is the case, we  expect to observe a
jump in  extinction from  one RC  to the other,  with each  RC getting
progressively  more reddened,  consistent with  the larger  distances.
This is  qualitatively what is observed,  judging by the CMD  shown in
Fig.~\ref{fig:cmd}.

The observed  mean magnitudes and colours  for these two RC  peaks are
$K_{\rm s},(J-K_{\rm  s})= 14.07, 1.52$ and  $K_{\rm s},(J-K_{\rm s})=
14.78, 1.62$, respectively. The inspection  of the control field shows
that only  the brighter  peak is  present.  In  order to  estimate the
distances  corresponding  to the  observed  magnitudes,  we adopt  the
intrinsic magnitudes and  colours of $M_{Ks} =-1.58 \pm  0.03$ and $(J
-Ks)  = 0.60  \pm 0.01$  for the  RC following  Alves et  al.  (2002).
These magnitudes  include the model correction  for population effects
(relative age and metallicity differences  from the Girardi \& Salaris
2001 models)  with respect to  the Large Magellanic Cloud  (LMC) where
$M_{Ks}=-1.60 \pm  0.03$.  This  is of  the same  order as  the offset
found by Pietrzynski  et al.  (2002) between local  dwarf galaxies and
the MW disk ($\Delta Ks_{HB}=0.025$  mag, their Fig.  7). Furthermore,
the mean values for the absolute magnitude of the RC in $K$~band given
by different authors agree to within  $<0.1$ mag (Alves 2000, Alves et
al.  2002, Pietrzynski et al.  2003, Grocholski \& Sarajedini 2002).

\begin{table}
\begin{center}
\caption{Measured magnitudes and colours of  the red clump peaks along
  the line of  sight and their derived extinction  and distances based
  on two different extinction laws.}
\begin{tabular}{lcc}
\hline
 & Brighter peak & Fainter peak \\
\hline
$K_{\rm s}$ [mag]                & 14.07$\pm$0.05  & 14.78$\pm$0.09 \\
$(J-K_{\rm  s})$ [mag]           & 1.52$\pm$0.04   &  1.62$\pm$0.14   \\
$E(J-K_{\rm  s})$ [mag]          & 0.92            &    1.02        \\
$A_K $ [mag] & 0.44$^a$ (0.67)$^b$      & 0.49$^a$ (0.74)$^b$   \\
$D $ [kpc]  &  11.0$\pm$2.4$^a$ (10.1)$^b$     & 14.8$\pm$3.6$^a$ (13.5)$^b$\\
\hline
\end{tabular}
\\
\small{($^a$)~$A_K = 0.484 \times E(J-K)$ (final adopted value)}
\small{($^b$)~$A_K = 0.725 \times E(J-K)$ (Cardelli et al. 1989)}
\end{center}
\end{table}

The  EJKS reddening  map in  Fig.~\ref{fig:map} traces  the integrated
extinction along  the line of  sight, which  corresponds to a  mean of
$A_{Ks}=0.46$ mag, and  $E(J-K_{\rm s})=0.95$ for VVV~WIN~1713$-$3939.
We  cannot assume  that this  reddening is  concentrated in  any place
along the  line of sight, although  it is probably more  severe within
the spiral  arms.  The reddening vector,  as measured from the  CMD of
tile  d075 (and  adjacent tiles),  appears to  follow a  reddening law
slope that is  shallower than the traditional Cardelli  et al.  (1989)
value $R_V=3.1$, shown  to be adequate for  higher Galactic latitudes.
Therefore, we  believe that distances  computed using the  steep slope
from  Cardelli et  al.   1989 ($A_{Ks}=0.725  \times E(J-K_{\rm  s})$)
would   be   underestimated.   We   use   the   method  described   in
Alonso-Garc\'ia et al.   (2017) to measure the slope  of the reddening
vector  directly  from  the   CMD  as  $A_{Ks}=(0.484\pm0.040)  \times
E(J-K_{\rm s})$.  The  slope is measured over all RC  stars in the CMD
and therefore  it is a  function only of the  specific two-dimensional
(2-D) location, but  does not provide dependence on  variations of the
extinction law with the distance along  the line of sight.  This value
is similar to $A_{Ks}=0.43 \times  E(J-K_{\rm s})$ measured toward the
Galactic  centre  in  Alonso-Garc\'ia   et  al.   (2017),  and  almost
identical  to  $A_{Ks}=0.49 \times  E(J-K_{\rm  s})$  from the  recent
independent extinction law  study by Majaess et  al.  (2016), obtained
from  RR Lyrae  variables  and two  separate  populations of  Cepheids
across  the entire  Galaxy.  It  is also  slightly shallower  than the
value $A_{Ks}=0.528 \times E(J-K_{\rm s})$ adopted by Nishiyama et al.
(2009).

The  RC luminosity  function  for the  window  (converted to  distance
modulus) is shown in the right panel of Fig.~\ref{fig:cmd}, along with
the  data  of the  comparison  field.   Adopting the  reddening  slope
measured from  the RCs  as $A_{Ks}=0.484  \times E(J-K_{\rm  s})$, the
distance moduli are computed as  $\mu=-5+5\,{\rm log}\,d({\rm pc})=
K_{\rm   s}-0.484\,(J-K_{\rm  s})+1.870$,   following  the   procedure
described by Minniti  et al.  (2011).  We applied a  simple Gaussian +
polynomial fit in order to obtain  the RC mean distances. The distance
moduli  are $\mu=15.20\pm0.47$  and $15.85\pm0.53$,  yielding mean  RC
distances   of   $D=$~11.0$\pm$2.4~kpc  and   14.8$\pm$3.6~kpc.    For
comparison,  the distances  adopting  the  relative extinctions  (from
Catelan et  al.  2011) with a  slope that follows the  Cardelli et al.
(1989)  reddening law  $A_{Ks}=0.725 \times  E(J-K_{\rm s})$  are also
presented in Table~1.

The mean  errors in the determination  of the mean magnitudes  for the
different features can be derived from the fits to the distribution of
distance    moduli    for    the     selected    red    clump    stars
(Fig.~\ref{fig:cmd}). Scaling these errors to the colours according to
the reddening slope leads to  typical colour uncertainties for the red
clumps of $\sigma (J-Ks) = 0.10$  mag.  We have computed the distances
using two different  reddening slopes, in order to  illustrate that in
general the major source of uncertainties  is due to the adoption of a
specific  reddening law,  and  not from  the  fits to  the  RC in  the
colour-magnitude  diagrams.   This is  why  it  is important  to  make
differential distance  estimates whenever  possible in  Galactic plane
studies.

\section{Mapping distant features in the Galactic disk} 
\label{sec:sec5}

The  low  reddening  of   VVV~WIN~1713$-$3939  provides  us  with  the
opportunity to investigate  the entire extension of the  disk.  Such a
study  is  indeed  difficult  elsewhere  because  of  the  very  large
extinction   at  low   latitude  regions,   which  introduces   severe
incompleteness  of  the  main  CMDs features.   The  bottom  panel  of
Fig.~\ref{fig:gaia}  illustrates that  the  number of  sources in  the
window is significantly higher for the relatively bluer passbands.

\begin{figure}
\centering
\includegraphics[scale=0.3]{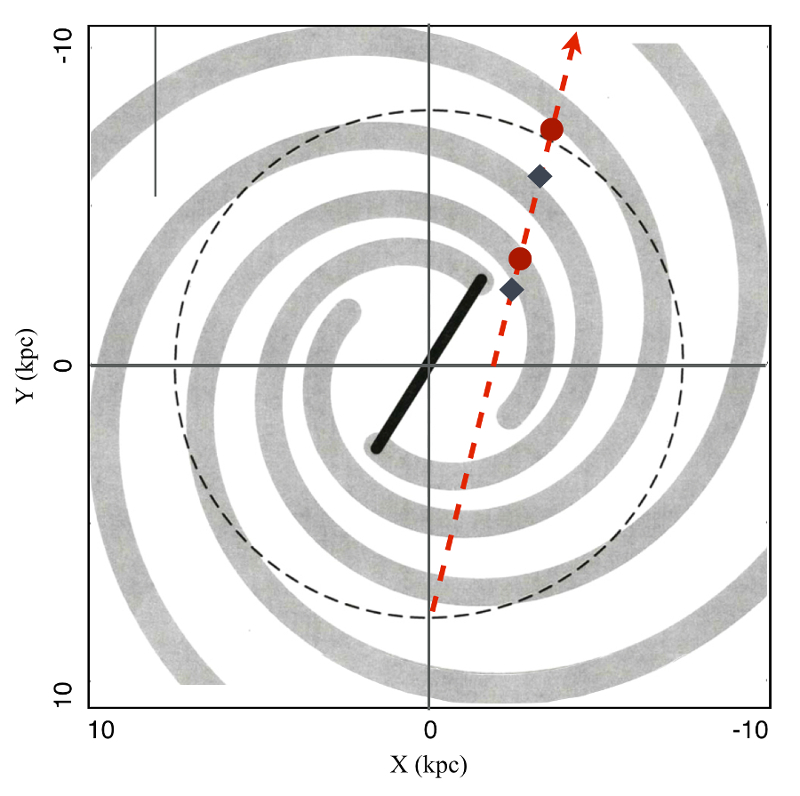}
\caption{Schematic  map  of  the  MW adapted  from  Vall\'{e}e  (2014)
  indicating the position of the Sun and the line of sight through the
  window.  We  adopted a  distance to the  Galactic centre  of $R_{\rm
    o}=8$~kpc. According  to this  picture, the  line of  sight should
  pass through  the Sagittarius arm and  the end of the  long bar, and
  the far  side of the  Norma and Scutum-Centaurus arms.   Red circles
  mark corresponding distances for  the main peaks using $A_{Ks}=0.484
  \times E(J-K_{\rm s})$.  Using  $A_{Ks}=0.725 \times E(J-K_{\rm s})$
  the distances are systematically shorter (black diamonds).}
\label{fig:mw}
\end{figure}

Figure~\ref{fig:mw} shows  the schematic  map of  the MW  adapted from
Vall\'{e}e (2014),  indicating the  line of sight towards  the window.
This picture  is in agreement with  the schematic face-on view  of our
Galaxy from Drimmel \& Spergel (2001) based on the analysis of near-IR
and far-IR data sets. The positions of the RC peaks are also marked.

The  most  important  conclusion  from this  comparison  is  that  our
measured  distances match  fairly  well with  those  in the  graphical
representation of the  Milky Way.  The closest RC, at  a mean distance
of  $D= 11.0$  kpc  (which is  also  seen in  the  control field),  is
consistent  with the  expected distance  to the  Galactic bar  at this
latitude (Wegg et  al.  2015).  This line of sight  intercepts the bar
at a projected distance of $D  \sim 3.5$ kpc from the Galactic centre,
approximately where the far side of the Sagittarius arm is expected to
start.

The second and faintest RC is at  $D=14.8$ kpc from the Sun and is not
seen   in  the   control  field   luminosity  function.    A  possible
interpretation that  follows from the comparison  in Fig.~\ref{fig:mw}
is that this  clump traces the location  of the far side  of the Norma
arm (or  even the Scutum-Centaurus arm)  along the line of  sight, not
detected in the control field because  of the high extinction. In such
a case, the line of sight  towards the window intercepts the near side
of the Norma  arm at a projected distance of  about $D=7$~kpc from the
Galactic centre.  Another possibility is that this peak corresponds to
the overall background disk  distribution, which would appear narrower
due to incompleteness at the faintest magnitudes.

Altogether, our  results suggest that VVV~WIN~1713$-$3939  is suitable
for studying  the overall structure  of the disk,  potentially tracing
important features (such  as spiral arms and the  general disk density
distribution) to constrain our current  representation of the far side
of  the  Galaxy.  Unfortunately,  the  choice  of the  extinction  law
shrinks or  expands the distance  scale, but we  argue that we  have a
qualitatively   good   agreement   with    the   Galactic   image   of
Figure~\ref{fig:mw}.   We  note that  the  Nishiyama  et al.   (2009),
Majaess   et   al.   (2016)   or   Alonso-Garc\'ia   et  al.    (2017)
selective-to-total extinction ratio laws  scale better to this picture
than steeper  options (e.g. Cardelli  et al. 1989, Rieke  \& Lebofsky
1985).

\section{Conclusions}

We have discovered VVV WIN 1713$-$3939, a low extinction window in the
inner plane of the Milky Way.  We show that it is possible to use this
window to chart  the innermost region of the  disk, placing structural
constraints on the  other side of the Galaxy. In  the past our ability
to perform such studies using stellar sources has been limited, if not
impossible, due to the large amounts of interstellar extinction in the
Galactic plane.  The new window has  a roughly circular shape of $30'$
diameter  size,  centred  at  $(l,  ~b) =  (347.4,  -0.4)$  deg.   The
extinction maps provide a  mean extinction of $A_{Ks}=0.46\pm0.04$ for
this   window,    corresponding   to   $E(J-K_{\rm    s})=0.95$,   and
$E(B-V)=1.45$.  This  is about  $\Delta K_{\rm s}  = 0.5$  and $\Delta
V=4.5$~mag  smaller  extinction,  on  average,  than  its  surrounding
fields.

We then  studied the stellar populations  along the line of sight
of this window (and a control  field) using the deep near-IR CMDs from
the  VVV Survey,  with the  aim of  estimating the  distances from  RC
giants, known  to be excellent  distance indicators. We  have computed
the  slope of  the  reddening law  towards  this line of sight,  which
appears  to  be  shallower  than   the  canonical  extinction  law  of
$R_V=3.1$.    We  find   the  slope   of  the   reddening  vector   is
$A_{Ks}=(0.484\pm  0.040) \times  E(J-K_{\rm s})$,  in agreement  with
Nishiyama (2009)  and more  recent studies  (Schultheis et  al.  2014,
Nataf et al.  2016, Majaess et al. 2016, Alonso-Garc\'ia 2017).

Finally, the  convenient location of this  window allows us to  gain a
view of  the far side  of our Galaxy.  Using  the CMDs we  can clearly
identify  two distant  RCs along  the  line of  sight.  These  stellar
over-densities are located  at mean distances of  $D= 11.0\pm2.4$ kpc,
and $14.8\pm3.6$ kpc  (adopting our measured extinction  law, which is
similar to that of Nishiyama et al.  2009).  These would correspond to
the intersection with  the Galactic bar and the far  Norma (or the far
Scutum-Centaurus) arm  on the opposite  side of the  MW, respectively.
An alternative  interpretation of  the most distant  RC peak,  that it
traces  the overall  background disk  population, cannot  be excluded.
Future  studies  including  proper  motion  measurements  and  Cepheid
variables to trace over-densities  among young stellar populations can
further probe the spiral structure of the Milky Way.

\begin{acknowledgements}

We gratefully acknowledge  the use of data from the  ESO Public Survey
programme  ID 179.B-2002  taken  with the  VISTA  telescope, and  data
products from  the Cambridge Astronomical Survey  Unit (CASU). Support
for the authors  is provided by the BASAL Center  for Astrophysics and
Associated Technologies (CATA) through  grant PFB-06, and the Ministry
for  the  Economy,  Development,   and  Tourism,  Programa  Iniciativa
Cienti\'ifica  Milenio   through  grant   IC120009,  awarded   to  the
Millennium  Institute  of  Astrophysics (MAS).   R.K.S.   acknowledges
support   from   CNPq/Brazil   through  projects   308968/2016-6   and
421687/2016-9.   R.K.  acknowledges  support from  CNPq/Brazil.  D.M.,
M.Z.,  R.B., and  J.A.G.   acknowledge support  from FONDECYT  Regular
grants  No.   1130196,  1140076,  and 1150345,  and  Iniciation  grant
No. 11150916, respectively.  We are  also grateful to the Aspen Center
for Physics  where our work  was partly supported by  National Science
Foundation  grant  PHY-1066293,  and  by   a  grant  from  the  Simons
Foundation (D.M.  and M.R.).

\end{acknowledgements}

\end{document}